\documentclass[prd,showpacs,preprintnumbers]{revtex4}

\usepackage{graphicx}
\usepackage{amsmath,amssymb}
\usepackage{epstopdf}

\newcommand\beq{\begin{equation}}
\newcommand\eeq{\end{equation}}
\newcommand\bea{\begin{eqnarray}}
\newcommand\eea{\end{eqnarray}}
\newcommand\bp{\boldsymbol{p}}
\newcommand\bq{\boldsymbol{q}}
\newcommand\bx{\boldsymbol{x}}

\begin{document}

\preprint{Imperial/TP/09/TK/02}

\title{Kinks and small-scale structure on cosmic strings}

\author{E.~J.~Copeland}
\email{ed.copeland@nottingham.ac.uk}
\affiliation{School of Physics and Astronomy, University of Nottingham, 
University Park, Nottingham NG7 2RD, United Kingdom}

\author{T.~W.~B.~Kibble}
\email{kibble@imperial.ac.uk}
\affiliation{Blackett Laboratory, Imperial College, London SW7 2AZ, United Kingdom}

\begin{abstract}
We discuss some hitherto puzzling features of the small-scale structure of cosmic strings.  We argue that kinks play a key role, and that an important quantity to study is their sharpness distribution. In particular we suggest that for \emph{very} small scales the two-point correlation function of the string tangent vector varies linearly with the separation and not as a fractional power, as proposed by Polchinski and Rocha [Phys.\ Rev.\  D {\bf 74}, 083504 (2006))].  However, our results are consistent with theirs, because the range of scales to which this linearity applies shrinks as evolution proceeds.
\end{abstract}


\maketitle

\section{Introduction}

From extensive studies over recent years a broad consensus has emerged about the evolution of a network of cosmic strings.  The various simulations all yield basically similar results \cite{Albrecht:1989mk,Bennett:1989yp,Allen:1990tv,Martins:2005es,Ringeval:2005kr,Urrestilla:2007yw,Hindmarsh:2008dw}.  The network will evolve towards a scaling regime in which the characteristic length scale $\xi$ grows in proportion to the age of the universe $t$.  The energy-loss mechanism that makes this possible is the formation and slow decay of loops.  However, many details, concerning small-scale structure, the typical sizes of the loops and the mechanism by which they lose energy and eventually disappear are still uncertain.  We do not know for sure whether the small-scale structure will also scale. 

The primary length scale $\xi$ of the network, the typical inter-string separation, may be defined by requiring that a randomly chosen large volume $V$ will on average contain a length of long string (\emph{i.e.}\ strings of length greater than the horizon size, excluding smaller loops)
 \beq 
 L = \frac{V}{\xi^2}. \label{L} 
 \eeq
Here, and subsequently, `length' of string means the \emph{invariant} length --- for a straight piece of string, the length in its rest frame.  Moreover, we use physical, not comoving, length units.  The assumption of scaling is that, after an initial transient period, $\xi$ will increase in proportion to the time.  Once scaling is achieved, the simulations tell us that
 \beq 
 \xi\approx\gamma t,\ \text{  with  }\ 
 \gamma_{\text{r}}\approx0.3, \quad
 \gamma_{\text{m}}\approx0.55, 
 \eeq
where the subscripts denote the radiation-dominated and matter-dominated eras.
 
From these simulations, it is hard to extract reliable information about small-scale structure, because we are talking about scales close to their resolution limit.  Moreover, the simulations can only cover a few Hubble times, while it is very possible that short-distance scaling will only be reached after a much longer time.  For that reason, analytic studies are important.

In our first attempt to construct an analytic model \cite{Kibble:1990ym} we introduced also a second length scale $\bar{\xi}$ defined in terms of the correlation function and representing the persistence length along the string, i.e., the distance beyond which there is no directional correlation.  Later this model was extended to the `three-scale model' of Austin \emph{et al} \cite{Austin:1993rg}, incorporating a length scale $\zeta$ loosely defined as a typical separation between kinks.  However, the treatment of small-scale structure relied on a number of untested assumptions and approximations and was not wholly satisfactory.  A considerable advance in a similar direction has been achieved more recently in the work of Polchinski and Rocha \cite{Polchinski:2006ee,Polchinski:2007rg}, based on a study of the correlation functions on strings.  They obtained very good agreement with results of simulations over the intermediate range of scales, but there remain puzzling features and unexplained discrepancies at very small scales.

In this note, we point out some significant issues, especially concerning the distribution of kinks on strings, and revisit some of the questions discussed in our earlier work \cite{Kibble:1990ym} which we believe may have an important bearing on the resolution of these problems.  We begin in Sec.~\ref{dynamics} by reviewing very briefly the dynamics of cosmic strings, and in Sec.~\ref{correlation} the estimation of correlation functions.  In Sec.~\ref{intercommuting} we discuss the calculation of the number of intercommuting events, and in Sec.~\ref{length} the derivation of the equation for the rate of change of the total length of long string.  Then in Sec.~\ref{kink} we study the distribution of kinks, which we argue holds the key to understanding the small-scale structure. The results are discussed in Sec.~\ref{discussion}.

\section{Dynamics}
\label{dynamics}

For simplicity, we consider only the standard (`vanilla') cosmic strings, moving in a spatially flat Friedmann-Lema\^{\i}tre-Robertson-Walker space-time, with metric
 \beq 
 ds^2 = dt^2 - a^2(t)d\bx^2, 
 \eeq
where $a\propto t^\nu$, with $\nu_{\text{r}}=\frac{1}{2}$ and $\nu_{\text{m}}=\frac{2}{3}$. 

The dynamics of cosmic strings on all scales much larger than the string width (and on time scales where gravitational or other radiation is insignificant) are well described by the Nambu-Goto equations.  They take their simplest form if we use null (characteristic) world-sheet coordinates $u$ and $v$, satisfying the conditions $x_{,u}^2=x_{,v}^2=0$, where the subscripts denote partial derivatives.  In flat space-time we can choose $u,v=t\pm s$ where $s$ measures the \emph{invariant} length along the string (the energy divided by the string tension $\mu$), and even in curved space-time we have the freedom to make reparametrizations $(u,v)\to (u'(u),v'(v))$, so we can always choose $u$ and $v$ \emph{locally} in this way.  When discussing small-scale features it is generally convenient to do so.  The Nambu-Goto equations then read
 \bea 
 t_{,uv}&=&-a^2 H(\bx_{,u}\cdot\bx_{,v}) \notag\\ 
 \bx_{,uv}&=&-H(t_{,u}\bx_{,v}+t_{,v}\bx_{,u}), \label{NG}
 \eea
where $H=\dot a/a$.  The characteristics are the constant-$u$ and constant-$v$ curves.

It is convenient to think of the string as a superposition of left- and right-moving components.  We introduce spatial unit vectors representing these waves, defined by 
 \beq 
 \bp=\frac{a\bx_{,u}}{t_{,u}}, \qquad  
 \bq=\frac{a\bx_{,v}}{t_{,v}}.
 \label{pq} 
 \eeq
Then the equations of motion (\ref{NG}) imply 
 \bea 
 \bp_{,v}&=&-Ht_{,v}(\bq-\bq\cdot\bp\bp), 
 \notag\\
 \bq_{,u}&=&-Ht_{,u}(\bp-\bp\cdot\bq\bq). 
 \eea
On time scales short compared to the Hubble time, the right-hand sides are small, so these equations show that $\bp$ varies only slowly along left-moving characteristics and $\bq$ along right-moving.  If we regard $\bp$ as a function of $u$ and the time $t$, and $\bq$ as a function of $v$ and $t$, these equations reduce to
 \bea 
 \dot\bp&=&-H(\bq-\bq\cdot\bp\bp), \notag\\
 \dot\bq&=&-H(\bp-\bp\cdot\bq\bq),
 \label{pqdot} 
 \eea
where the dots denote derivatives with respect to $t$.  In effect, the $\bp$ and $\bq$ vectors repel each other slightly on the unit sphere as they pass.

When strings meet they normally exchange partners, or `intercommute'.  This process creates on each of the two strings a pair of kinks moving in opposite directions.  One consequence of the stretching of strings by the universal expansion is a gradual diminution of the kink angle $\theta$.  At a left-moving kink the value of $\bp$ changes abruptly, say from $\bp_1$ to $\bp_2$.  Let us define the sharpness of the kink to be 
 \beq 
 \psi = \tfrac{1}{2}(1-\bp_1\cdot\bp_2) = \sin^2(\theta/2), 
 \eeq
which of course ranges from 0 to 1.  Then from (\ref{pqdot}) we have
 \beq 
 \dot\psi = \tfrac{1}{2} H(\bq\cdot\bp_2
 - \bq\cdot\bp_1\, \bp_1\cdot\bp_2
 + \bp_1\cdot\bq - \bp_1\cdot\bp_2\, \bp_2\cdot\bq). 
 \label{dotpsi} 
 \eeq
Since the time scale for change of $\psi$ is the Hubble time, we can replace the scalar products with $\bq$ by their time average, related to the mean square velocity by
 \beq 
 \langle \bp\cdot\bq \rangle \equiv - \bar\alpha 
 = -(1-2\langle \dot\bx^2 \rangle). 
 \label{baral}
 \eeq
According to the simulations,
 \beq
 \bar\alpha_{\text{r}} \approx 0.18, \qquad 
 \bar\alpha_{\text{m}} \approx 0.3.
 \eeq
Thus (\ref{dotpsi}) becomes
 \beq 
 \dot\psi = -2H\bar\alpha\psi, 
 \eeq
whence it follows at once that 
 \beq
 \psi \propto a^{-2\bar\alpha} \propto t^{-2\zeta},
 \label{blunting}
 \eeq
where
 \beq 
 \zeta = \nu\bar\alpha, \quad\text{or} \quad
 \zeta_{\text{r}} \approx 0.09, \quad \zeta_{\text{m}} 
 \approx 0.2.
 \eeq
Note that in the radiation-dominated era this blunting of kinks is an exceedingly slow process: for the sharpness to reach half its value at $t=1$ we must wait until $t=47$ (though in the matter era, $t=5.7$ would suffice).

\section{Correlation functions}
\label{correlation}

Polchinski and Rocha \cite{Polchinski:2006ee} used these same equations to derive expressions for the $\bp$-$\bp$ (or $\bq$-$\bq$) correlation function on the strings.  In particular, they argued that the correlation function should have the form
 \beq 
 z(s,t) \equiv \langle\bp(u_1,t)\cdot\bp(u_2,t)\rangle 
 = 1-A(s/t)^{2\chi},
 \label{pcorr} 
 \eeq
with $s=u_1-u_2$, where we have chosen the null coordinates locally to be $u,v = t\pm s$.  Here $A$ is a constant and 
 \beq
 \chi = \frac{\zeta}{1-\zeta}, 
 \qquad\text{or} \qquad
 \chi_{\text{r}} \approx 0.10, \quad 
 \chi_{\text{m}} \approx 0.25. 
 \eeq
This expression provides a very good fit to the data from simulations over a considerable range of scales, but at very small scales, particularly in the radiation era, the correlation function seems to approach unity much faster than this would suggest.  In fact, we shall argue that (\ref{pcorr}) \emph{cannot} be correct at very small scales.  Nevertheless, as we shall see, it may still be true in some limiting sense.

At the end of the friction-dominated era, the strings are expected to be smooth, with no sharp kinks.  In the subsequent evolution, they will always remain smooth except for the kinks introduced by intercommuting events.  But in a finite time there can only be a finite number of such events --- later we shall estimate the number --- so the strings should always be piecewise smooth.  This means that on any short stretch of string of length $s$ the probability of finding a kink goes to zero in proportion to $s$.  It follows that for very small $s$, $1-z(s,t)\propto s$; it \emph{cannot} behave like any fractional power of $s$.  The kink distribution is, we contend, the key to understanding the very small scale properties of the string network.

\begin{figure}
 \begin{center}
  \includegraphics[width=3in,angle=0]{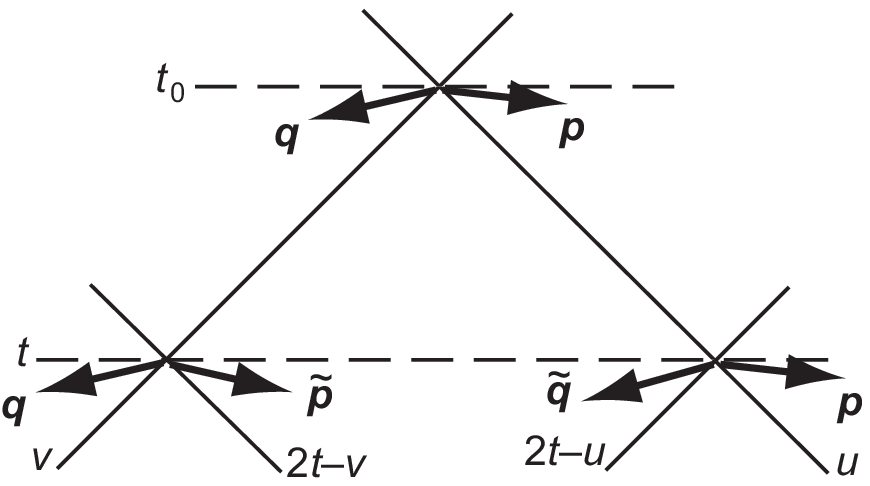}
 \caption{Vectors involved in the rate of change of $\bp\cdot\bq$.}
 \label{pqcor}
 \end{center}
 \end{figure}

As discussed in ref.~\cite{Kibble:1990ym} the equations (\ref{pqdot}) can also be used to relate the correlation function $z$ to $\bar\alpha$ (though some of the factors in that earlier discussion were incorrect).  The vectors $\bp(u,t)$ and $\bq(v,t)$ meet when $2t=u+v=2t_0$, say.  The only reason they are then correlated, albeit weakly, is of course the `repulsion' effect.  As it approaches $\bp$ our chosen vector $\bq$ is `repelled' by other $\bp(u')$ vectors, and although these small changes fluctuate in direction, so long as the general direction is the same, there will be a net effect.  We have
 \beq
 \partial_t (\bp\cdot\bq) \approx
 - H[(\bp\cdot\tilde\bp - \bp\cdot\bq\,\bq\cdot\tilde\bp) 
 + (\tilde\bq\cdot\bq - \tilde\bq\cdot\bp\,\bp\cdot\bq)],
 \eeq
where (see Fig.~\ref{pqcor}) $\bp$ and $\bq$ stand for $\bp(u,t)$  
and $\bq(v,t)$ while
 \beq
 \tilde\bp=\bp(2t-v,t),\qquad \tilde\bq=\bq(2t-u,t).
 \eeq
Since the correlation between the $\bp$ and $\bq$ vectors is very weak, the ensemble average of the expression in the first parenthesis is to a good approximation $\frac{2}{3}\langle \bp\cdot\tilde\bp \rangle$, and similarly for the second parenthesis.  Thus we find
 \beq
 \partial_t \langle \bp(u,t) \cdot\bq(v,t) \rangle \approx 
 - \frac{4}{3} H(t)z(u+v-2t,t),
 \eeq
which must then be integrated up to the time $t_0$ from an early time $t_1$ before which the correlation was negligble:
 \beq
 \langle \bp(u,t) \cdot\bq(v,t) \rangle \approx 
 -\frac{4}{3} \int_{t_1}^{t_0} dt'\,H(t')z(2t_0-2t',t').
 \eeq
The largest contribution comes from the region where $2t_0-2t' \ll t'$, so to a first approximation we can neglect the $t'$ dependence of $H$ and the explicit $t'$ dependence of $z$, obtaining
 \beq
 \bar\alpha \equiv -\langle \bp\cdot\bq \rangle
 \approx \frac{2}{3} H(t)\bar\xi(t),
 \eeq 
where $\bar\xi$ is the second length scale introduced in ref.~\cite{Kibble:1990ym}, defined by
 \beq
 \bar\xi(t) = \int_0^\infty z(s,t)\,ds.
 \eeq
Note that the upper limit has been extended to infinity on the assumption that $z$ vanishes for $s>2(t_0-t_1)$.  This $\bar\xi$ represents the persistence length along the (left-moving) string --- the distance it is expected to go on traveling in the same direction.

This second length should also scale:
 \beq \bar\xi = \bar\gamma t, \eeq
and we then have the relation
 \beq \bar\alpha \approx \frac{2\nu}{3}\bar\gamma. \eeq
This would yield
 \beq \bar\gamma_{\text{r}} \approx 0.5,\qquad
 \bar\gamma_{\text{m}} \approx 0.7.
 \eeq
Interestingly, this suggests that in both cases, $\bar\xi$ is slightly larger than $\xi$.  Note that although $\bar\xi$ is similar to the correlation length used among others by ref.~\cite{Martins:2005es} (there called $\xi$) it is not identical. 

\section{Intercommuting}
\label{intercommuting}

Next, we review the calculation of the rate of intercommuting.  We consider a large volume $V$ and ask how many such events occur within a short time interval $dt$.  The total length of string within $V$ is $L$, given by (\ref{L}).  Within this thin slice, we can choose the world-sheet coordinates to be $u,v=t\pm s$, so that $dtds=\frac{1}{2}dudv$.  Now let us randomly select two small sections of the string world-sheet, say with coordinates in the ranges $u_j$ to $u_j+du_j$ and $v_j$ to $v_j+dv_j$ ($j=1,2$).  For the moment, we ignore the special case of direct small-loop creation, and assume that these sections are uncorrelated.  The probability that the two will intersect is then the four-volume $d\Omega$ that they span divided by the total four-volume $Vdt$.  Evidently,
 \beq 
 d\Omega = |\sqrt{-g}\epsilon_{\lambda\mu\nu\rho}
 x^\lambda_{1,u}x^\mu_{1,v}
 x^\nu_{2,u}x^\rho_{2,v}|\,du_1dv_1du_2dv_2. \eeq
We can then extract factors of $t_{,u}=t_{,v}=\frac{1}{2}$ and rewrite this determinant in terms of the $\bp$ and $\bq$ vectors defined in (\ref{pq}), obtaining
 \beq 
 d\Omega = \Delta.\frac{1}{2}du_1dv_1.\frac{1}{2}du_2dv_2, 
 \eeq
where $\Delta$ is a $4\times 4$ determinant,
 \bea 
 \Delta &=& \frac{1}{4}\Bigg| \begin{vmatrix} 1 & 1 & 1 & 1 \\
 \bp_1 & \bq_1 & \bp_2 & \bq_2 \end{vmatrix} \Bigg| \notag\\
 &=& \frac{1}{4}\big|(\bp_1 \times \bp_2)\cdot (\bq_1-\bq_2)
 +(\bp_1- \bp_2)\cdot (\bq_1 \times \bq_2)\big|.
 \label{Delta}
 \eea

The probability $dp_1$ that the first section of the world sheet will experience an intercommuting with \emph{any} other section is obtained by integrating over all $u_2,v_2$, thus replacing $\frac{1}{2}du_2dv_2$ by $Ldt$:
 \beq dp_1 =  \frac{\bar\Delta}{\xi^2}\frac{1}{2}du_1dv_1, \eeq
in which we have used eq.~(\ref{L}), and where $\bar\Delta$ is the average value of $\Delta$. Thus finally the total number of intercommmuting events in the volume $V$ within the time interval $dt$ is
 \beq dN_{\text{intercom}} = \frac{\bar\Delta Vdt}{2\xi^4}. \eeq
The factor of 2 here is to compensate for the double counting of events.

The average value of $\Delta$ was computed in ref.~\cite{Austin:1993rg}, eq.~(4.30) (there called $\chi$), assuming that the unit vectors are independently uniformly distributed over the unit sphere:
 \beq 
 \bar\Delta =  \int \frac{d^2\bp_1}{4\pi}\frac{d^2\bq_1}{4\pi}
 \frac{d^2\bp_2}{4\pi}\frac{d^2\bq_2}{4\pi} \Delta,
 \label{barDelta}
 \eeq  
where each $d^2\bp$ signifies an integral over the unit sphere. The result is 
 \beq
 \bar\Delta = \frac{2\pi}{35} \approx 0.18.
 \eeq
We can allow for the small $\bp$-$\bq$ anticorrelation in (\ref{baral}), by assuming a linear probability distribution for $\bp\cdot\bq$ and inserting in the integrand of (\ref{barDelta}) the two factors $(1 - 3\bar\alpha \bp_j \cdot\bq_j)$.  This increases $\bar\Delta$ marginally to
 \beq
 \bar\Delta = \frac{2\pi}{35}
 \left(1+\frac{2\bar\alpha}{3}-\frac{\bar\alpha^2}{11}\right),
 \eeq
which would yield
 \beq
 \bar\Delta_{\text{r}}\approx 0.20,\qquad
 \bar\Delta_{\text{m}}\approx 0.21.
 \eeq

One interesting feature of $\Delta$, which will be important later, should be noted.  It vanishes when any pair of the four unit vectors $\bp_1,\bp_2,\bq_1,\bq_2$ is equal, but \emph{not} when they are equal and opposite.  This has an important implication for kink formation.  Because the probability of intercommuting is zero when $\bp_1=\bp_2$ but not when $\bp_1=-\bp_2$, the angle distribution of newly formed kinks (computed below) is not symmetrical but is actually skewed towards sharper kinks.  It appears that $\Delta$ attains its maximum value, $4/(3\surd3)=0.77$, when the four unit vectors are aligned towards the vertices of a regular tetrahedron, with equal angles, $\arccos(-\frac{1}{3})$, between each pair.    

We have so far ignored the direct formation of very small loops, which is really a separate type of process.  In that case, there is a high degree of alignment between all the $\bp$ and $\bq$ vectors, so the value of $\Delta$ is always small, nowhere near its average value.  Then the kinks produced are normally small-angle.

\section{Rate of change of length}
\label{length}

Let us now consider the rate of change of the total length $L$ of long string in the comoving region of volume $V$.  It will change for two main reasons: stretching by the universal expansion, and excision of portions of string by loop formation.

The rate of change due to stretching is a simple function of the r.m.s.\ velocity and the expansion rate \cite{Polchinski:2006ee}, namely
 \beq 
 \left(\frac{\dot L}{L}\right)_{\text{stretching}} = 
 \bar\alpha H,
 \eeq
where $\bar\alpha$ is given by (\ref{baral}). 

Next we ask how many of the intercommmutings will generate loops.  We expect that in the scaling regime a small, but fixed, fraction of the intercommutings will yield loops of substantial size.  Some of those will subsequently reconnect to the long-string network.  Many others will self-intersect and fragment, leading after a few oscillation periods to a final population of much smaller loops that oscillate and very gradually shrink.

Each of the initial loops will remove a length of string that is on average some fraction of $\xi$.  Hence the total length removed within a time interval $dt$ by loops that do not reconnect will be $dL=-\eta V/\xi^3$ for some fixed $\eta$, which must be determined from simulations.  (This parameter $\eta$ is related to the parameter $qc$ introduced in eq.~(4.31) of ref.~\cite{Kibble:1990ym}.) 

This then gives 
 \beq 
 \left(\frac{\dot L}{L}\right)_{\text{loop\ form}} = 
 -\frac{\eta}{\xi} = -\frac{\eta}{\gamma t}.
 \label{loopform}
 \eeq
 
Adding these two contributions the total rate of change of $L$ will be
 \beq 
 \frac{\dot L}{L} = \frac{\zeta}{t}-\frac{\eta}{\gamma t}.
 \eeq
Using (\ref{L}) this can also be written in terms of $\xi$ or $\gamma=\xi/t$, as
 \beq 2t\dot\gamma = \eta - (2-3\nu+\zeta)\gamma. 
 \eeq

This confirms that $\gamma$ will approach a constant scaling value, at which the right hand side vanishes.  Using the values from the simulations this gives
 \beq 
 \eta_{\text{r}}\approx 0.18, \qquad 
 \eta_{\text{m}}\approx 0.1. 
 \eeq

\section{Kink distribution}
\label{kink}

Here we consider the number of kinks within the volume $V$ and their sharpness distribution.  Each intercommuting creates two new left-moving and two right-moving kinks.  We expect the numbers to remain equal on average, though excised loops may by chance contain more of one than the other.  So let us consider the total number $N(\psi,t)d\psi$ of left-moving kinks with sharpness between $\psi$ and $\psi+d\psi$.

Kinks are created by intercommuting, with some initial sharpness distribution $g(\psi)$.  If we ignore kinks created when small loops are formed (for which the values of $\bp_1$ and $\bp_2$ are strongly correlated), then this distribution can be found by averaging the function $\Delta$ of (\ref{Delta}) over all directions of $\bq_1$ and $\bq_2$.  (This can most easily be done by transforming to the variables $\frac{1}{2}(\bp_1\pm\bp_2)$ and $\frac{1}{2}(\bq_1\pm\bq_2)$, and choosing the $z$ axis in the direction of $\frac{1}{2}(\bp_1-\bp_2)$, and $\frac{1}{2}(\bq_1-\bq_2)$ in the $xz$ plane.)  For simplicity, we ignore here the effect of the small anti-correlation of $\bp$ and $\bq$, and assume that the $\bq$ vectors are uniformly distributed on the unit sphere.  The result is
 \beq
 g(\psi) = \frac{35}{256}\sqrt{\psi}(15-6\psi-\psi^2).
 \label{initkinks}
 \eeq
\begin{figure}
\begin{center}
  \includegraphics[width=3in,angle=0]{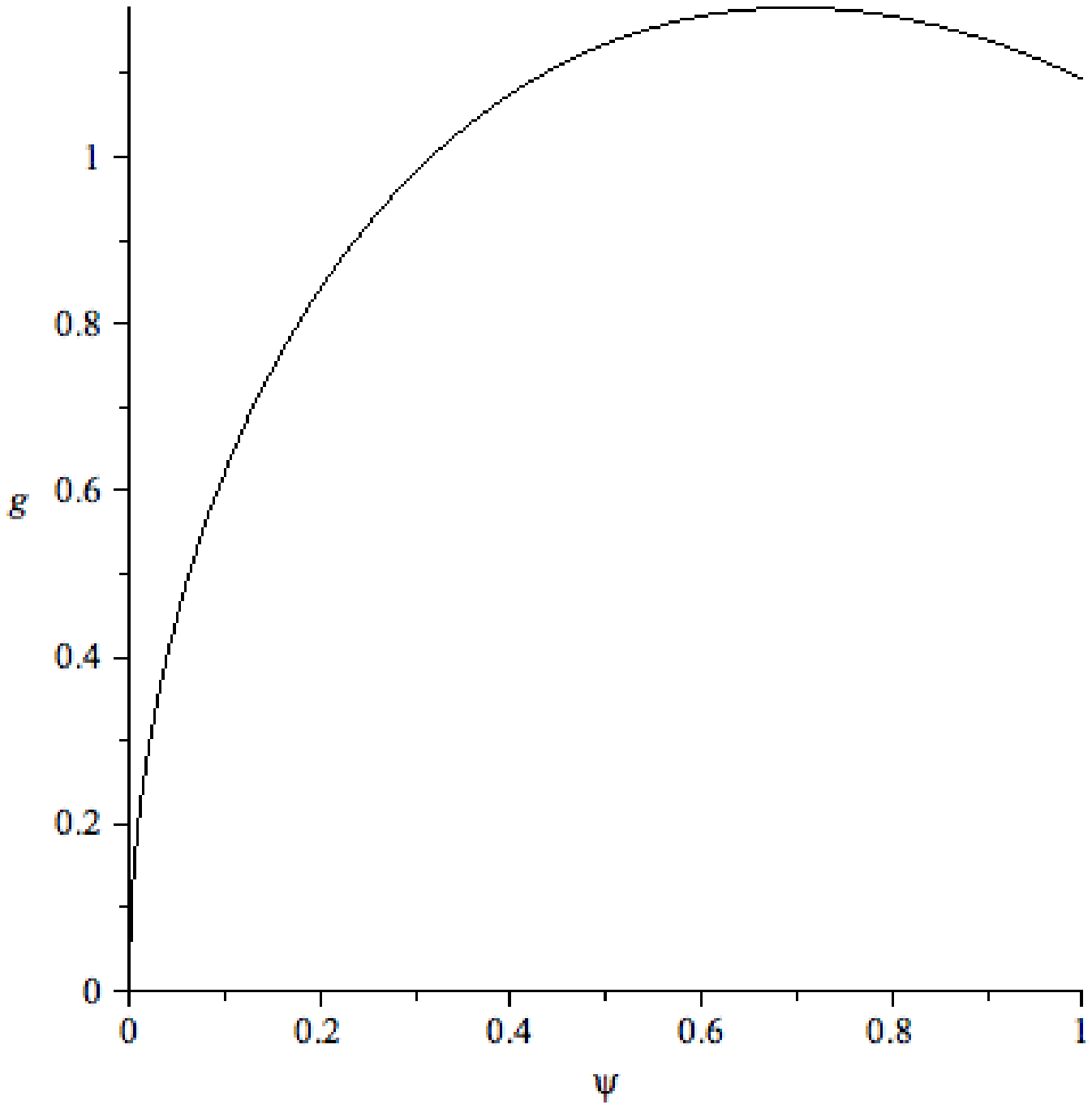}
\caption{The initial kink sharpness distribution}
\label{gpsi}
\end{center}
\end{figure}
This function is plotted in Fig.~\ref{gpsi}.  The mean value of $\psi$ is easily found to be 
 \beq
 \bar\psi = \int_0^1 \psi g(\psi)d\psi = \frac{5}{9}.
 \label{psibar}
 \eeq
As expected from our earlier discussion, this is slightly larger than $\frac{1}{2}$.  Of course, allowing for the kinks produced when small loops are formed would probably lower the average. 

Once formed, the kinks weaken, according to (\ref{blunting}).  Some of them are of course removed by being incorporated into loops.  A reasonable first guess might be that  the fraction of kinks removed is approximately the same as the fraction of total length excised.  However, there may be some tendency for the regions incorporated into loops to have more kinks than others, so to allow for that we shall assume that in place of (\ref{loopform}) we have
 \beq 
 \left(\frac{\dot N}{N}\right)_{\text{loop\ form}} = 
 -\frac{\bar\eta}{\xi} = -\frac{\bar\eta}{\gamma t},
 \eeq
where $\bar\eta$ is another parameter, possibly somewhat larger than $\eta$.  This should again be determined from simulations.

Putting these three effects --- kink formation by intercommuting events, blunting due to stretching, and removal by loop formation --- together we then get 
 \beq 
 \dot N = \frac{\bar\Delta V}{\gamma^4 t^4}g(\psi) + 
 \frac{2\zeta}{t}\frac{\partial}{\partial\psi}(\psi N) - 
 \frac{\bar\eta}{\gamma t}N.
 \eeq

To solve the equation we also need an initial condition.  In the radiation-dominated era, the proper condition is that at the end of friction domination there are no kinks at all.  Let us suppose that that occurs at a time $t_*$.  It is useful to change variable from $\psi$ to $\psi_* = (t/t_*)^{2\zeta}\psi$, which is the value $\psi$ would have had at $t_*$.  This yields the equation
 \beq
 t\dot N(\psi_*,t) + 
 \left(\frac{\bar\eta}{\gamma}-2\zeta\right)N(\psi_*,t) = 
 \frac{\bar\Delta V}{\gamma^4 t^3}
 g\left(\left(\frac{t_*}{t}\right)^{2\zeta} \psi_*\right),
 \eeq
where $\dot N$ on the left hand side now denotes the time derivative at constant $\psi_*$.  Note that $\psi_*$ may be larger than unity.  In that case, we need the additional boundary condition that $g(\psi)=0$ for $\psi>1$.  Then it is straightforward to solve the equation and return to $\psi$ rather than $\psi_*$.  Assuming that scaling has already been reached, so that $\dot\gamma=0$, the result is  
 \beq
 \frac{N(\psi,t)}{V} = 
 \frac{\bar\Delta}{\gamma^4 t^{3-\beta}} 
 \int_{\max(t_*,\psi^{1/2\zeta}t)}^t 
 \frac{dt'}{{t'}^{1+\beta}}\,
 g\left(\left(\frac{t}{t'}\right)^{2\zeta} \psi\right),
 \eeq
where
 \beq 
 \beta = 3-3\nu-\frac{\bar\eta}{\gamma}+2\zeta. 
 \eeq
If we ignore the possibility that loops have more kinks per unit length than average, and set $\bar\eta=\eta$, this would imply
 \beq 
 \beta_{\text{r}}\approx 1.1, \qquad 
 \beta_{\text{m}}\approx 1.2,
 \eeq
so we may perhaps expect that the true values are a little smaller than these.  Thus we obtain two separate expressions for $N$ depending on whether $\psi$ is larger or smaller than $(t_*/t)^{2\zeta}$. 

If we again ignore the kinks generated by small-loop formation, and use (\ref{initkinks}) for the initial kink sharpness distribution, then it is easy to compute the distribution at a later time.  It is convenient to write $g(\psi)$ as 
 \beq
 g(\psi)=\sum_k g_k \psi^k,
 \eeq 
where $k$ runs over the values $k=\frac{1}{2},\frac{3}{2},\frac{5}{2}$, and the $g_k$ can be read off from (\ref{initkinks}).  We then obtain
 \bea
 \frac{t^3 N(t,\psi)}{V} &=& \sum_k
 \frac{g_k\bar\Delta}{(\beta+2k\zeta)\gamma^4}
 (\psi^{-\beta/2\zeta}-\psi^k),
 \qquad \psi>(t_*/t)^{2\zeta} \label{kinkscaling}\\
 \frac{t^3 N(t,\psi)}{V} &=& \sum_k
 \frac{g_k\bar\Delta\psi^k}{(\beta+2k\zeta)\gamma^4}
 \left[\left(\frac{t}{t_*}\right)^{\beta+2k\zeta}
 - 1\right], \qquad \psi<(t_*/t)^{2\zeta}. 
 \eea

For large angles, this is a scaling distribution --- the right hand side of (\ref{kinkscaling}) is time-independent.  But the small-angle distribution does not scale.  On the other hand, the limiting value $\psi_{\text{c}}=(t_*/t)^{2\zeta}$ above which scaling holds decreases, albeit slowly, with time.
   
These expressions are necessarily limited to the radiation-dominated era.  In the matter-dominated era, we should use instead the sharpness distribution at the end of the radiation era as initial condition.  We shall not pursue that calculation here, but for observational predictions it may be necessary to do so. However, it should be noted that by the onset of the matter era, the limiting value $\psi_{\text{c}}$ is already extremely small. 

\section{Discussion}
\label{discussion}

The results obtained have some important implications both for observational predictions and for comparisons with simulations.

At times of any interest to observational predictions, the critical value $\psi_c$ is exceedingly small, so over almost the entire range of $\psi$ we have the solution (\ref{kinkscaling}).  This is confirmation that the kink distribution over most of its range does reach a scaling regime.  On the other hand, when it comes to comparisons with simulations, $\psi_c$ will still be quite large, so it is clear that in this particular respect we cannot expect the results to represent the full scaling regime.

Note also that the number of kinks increases towards $\psi=0$.  This is because of the large number of kinks produced at much earlier times that have gradually become less sharp.  The total number of kinks at any time is of course finite, though if (\ref{kinkscaling}) were valid down to $\psi=0$ it would not be so.  In fact the number of kinks per unit length, in units of $t$ is 
 \beq
 \frac{tN_{\text{kink}}}{L} = \frac{t}{L}\int_0^1 N(\psi,t)d\psi
 = \frac{\bar\Delta}{\gamma^2(\beta-2\zeta)}
 \left[\left(\frac{t}{t_*}\right)^{\beta-2\zeta}-1\right],
 \label{kinkno}
 \eeq
which increases quite rapidly with time.  If we use the values of the parameters estimated from simulations, we find in the radiation era
 \beq
 \frac{tN_{\text{kink}}}{L} 
 \approx 2.4 \left[\left(\frac{t}{t_*}\right)^{0.9}-1\right].
 \eeq
 
Note, however, that the large number of kinks mostly have very small sharpness.  Taking kinks as a whole, the mean inter-kink distance $L/N_{\text{kink}}$ does not scale.  However, if we choose some fixed lower limit to the sharpness, the average distance between the kinks above that limit \emph{does} eventually scale.  The density of kinks grows with time, but the excess is entirely in very small-angle kinks.

We can also now return to the question of the correlation function $1-z(s)=1-\langle\bp(0)\cdot\bp(s)\rangle$ on the strings.  On very small scales, the \emph{only} contribution comes from kinks.  A kink of sharpness $\psi$ between $0$ and $s$ will contribute an amount $2\psi$, and if $s$ is small enough that no more than one kink is likely to be found on the segment, then one simply has to integrate over the kink probability distribution, obtaining
 \beq
 1-z(s,t) = \int_0^1 \frac{sN(\psi,t)}{L}2\psi d\psi.
 \eeq
Evaluating the integral explicitly gives a very similar expression to (\ref{kinkno}), namely
 \beq
 1 - z(s,t) = 
 \frac{2s}{t} \frac{\bar\psi\bar\Delta}
 {\gamma^2(\beta-4\zeta)}
 \left[\left(\frac{t}{t_*}\right)^{\beta-4\zeta}-1\right],
 \eeq
where $\bar\psi$ is the average sharpness of newly formed kinks, given by (\ref{psibar}), namely $\frac{5}{9}$.  Again using the earlier estimates of the parameters yields
 \beq
 1 - z(s,t) \approx 
 3.3 \frac{s}{t} \left[\left(\frac{t}{t_*}\right)^{0.7}-1\right].
 \label{kinkcorr}
 \eeq
 
By comparison the power-law expression (\ref{pcorr}) \cite{Polchinski:2006ee}, gives
 \beq
 1 - z_{\text{PR}}(s,t) \approx 
 0.6\left(\frac{s}{t}\right)^{0.2}.
 \label{zPR}
 \eeq

As expected, the expression (\ref{kinkcorr}) is proportional to $s$, not to a fractional power of $s$.  However it is \emph{not} simply proportional to $s/t$, so it does not scale: the factor in brackets increases with $t$, and would become infinite in the limit $t\to\infty$.  This calculation makes sense only for small enough values of $s/t$ that there is unlikely to be more than one kink present.  As the number density of kinks, in terms of the scaling variable $s/t$, increases, the range of values of $s/t$ decreases.  To extend the calculation directly to larger values, we would need to know not just the density of kinks, but the extent of the correlation between the orientations of nearby kinks.

Thus the picture we are suggesting is that at any time, the correlation function has a linear behavior for very small $s/t$, and that as a function of $s/t$ it becomes steeper as time goes on.  At the same time the range of values of $s/t$ over which this linear behavior is expected decreases.  For larger values of $s/t$ we expect that the power-law behavior of (\ref{zPR}) will apply.  It is interesting to note that the point $s_{\text{c}}$ at which the the two curves (\ref{kinkcorr}) and (\ref{zPR}) cross, and which may be taken as a rough estimate of where one type of behavior changes to the other, is for large times given by 
 \beq
 \frac{s_{\text{c}}}{t} \approx 
 0.12 \left(\frac{t_*}{t}\right)^{0.9}.
 \label{stcrit}
 \eeq
In physical units, this limiting value $s$ apparently increases with time, but only very slowly, like $t^{0.1}$ (though it should be noted that the error in this exponent may well be of order $0.1$).

\begin{figure}
\begin{center}
  \includegraphics[width=3in,angle=0]{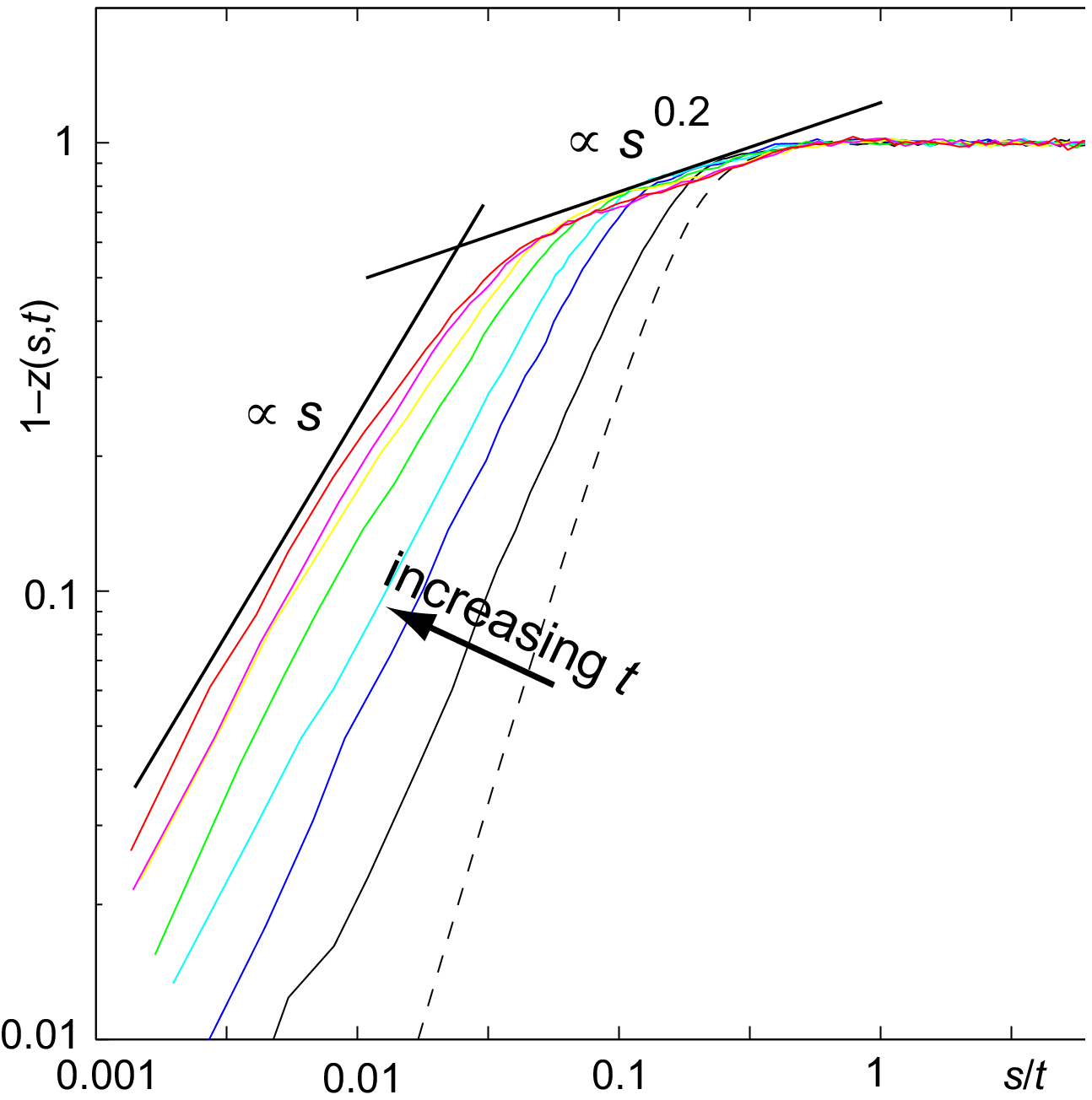}
\caption{Values of the correlation function $1-z(s,t)$ as a function of $s/t$.  Successive curves from right to left are at times $t/t_*=1$ (dashed), $1.77$ (black), $2.76$ (blue), $3.96$ (cyan), $5.43$ (green), $7.08$ (yellow), $8.94$ (magenta), $10.0$ (red).}
\label{MSsim}
\end{center}
\end{figure}
These predictions can be compared with the simulations of \cite{Martins:2005es} and, as shown in Fig.~\ref{MSsim}, they fit very well.  The figure shows a log-log plot of $1-z(s,t)$ as a function of $s/t$ for eight different times, with values of $t$ increasing from right to left.  All the curves follow the Polchinski-Rocha form (\ref{zPR}) roughly from $s/t\sim 1$ down to a limiting value that moves down as $t$ increases.  For smaller values of $s/t$ the curves steepen fairly sharply and at least for the later times fit well with the linear form (\ref{kinkcorr}), and as expected the turnover point moves leftwards with increasing time.  Naturally, the transition between the two regimes is not sharp, but nevertheless very clear. Given that the simulations start from an initial state in which the strings are smooth, it is appropriate to set $t_*$ equal to the initial time.  Then the latest time, represented by the leftmost (red) curve is at $t/t_*= 10.0$.  Of course the coefficient of $s/t$ in (\ref{kinkcorr}) at the initial time vanishes, so there we do not expect linear behavior.  For the next one or two times, the curves are still somewhat steeper than linear, but thereafter (\ref{kinkcorr}) gives a good fit.  In checking the rate at which the turnover value of $s/t$ decreases, it should be noted that for the earlier times we cannot use (\ref{stcrit}), because the term 1 in (\ref{kinkcorr}) cannot be neglected.  At least for the later curves, the rate appears entirely consistent.
 
At times of observational interest, we believe the Polchinski-Rocha formulae of ref.~\cite{Polchinski:2006ee} should hold on most relevant scales, though it is conceivable that the different behavior at very small scales might be relevant in some cases.  It seems plausible that the reason for the discrepancy at small $s$ between their results and the simulations is that while the simulations correctly exhibit scaling on large and indeed intermediate scales, they have not yet reached the point of complete scaling at very small scales.  It should, however, be possible to check from simulations whether this prediction is correct.

It should be noted that one thing we have not considered at all is the effect of gravitational or other radiation.  This could change the picture, but only by making the strings smoother, not less smooth, so it could not affect our main conclusions.

\begin{acknowledgments}

We are indebted to Carlos Martins and Paul Shellard for providing us with results from their simulations.  We thank Joe Polchinski for a valuable discussion.

\end{acknowledgments}

\end{document}